\RequirePackage[2020-02-02]{latexrelease}
\documentclass[twocolumn,prl,amsmath,amssymb,showpacs,superscriptaddress,floatfix]{revtex4}
\usepackage{graphicx}
\usepackage{bm}
\usepackage{lineno,hyperref}
\usepackage{epsf}
\usepackage{xcolor}
\begin{document}
\title{The origin of complex behavior of liquid carbon: an insight from computer simulation}

\author{Yu. D. Fomin \footnote{Corresponding author: fomin314@mail.ru}}
\affiliation{Vereshchagin Institute of High Pressure Physics,
Russian Academy of Sciences, Kaluzhskoe shosse, 14, Troitsk,
Moscow, 108840, Russia }
%

%

%
\date{\today}

%

\begin{abstract}

In the present paper we perfomrm molecular dynamics simulation of liquid carbon with 
a machine-learning potential GAP-20. We show that within the framework of this model
carbon demonstrates a relatively low critical temperature, which can
affect the results of experimental measurements of melting point of
graphite.

\end{abstract}


\maketitle


\section{Introduction}

Melting of graphite is one of a long-standing problems of physics. Although the first experiments
were undertaken in the beginning of XX century \cite{1911} (see, for instance,
Chapter 1 of the book \cite{savvatimsky} for a historical review),
the exact location of the melting point is still controversial. In spite of numerous works in this field
different experiments give the estimation
of melting point of graphite from $T_m=4000$ K \cite{savvatimsky} up to $T_m=6700$ K \cite{rakhel}.
Apparently, such dispersion of the results cannot be recognized as satisfactory. Therefore, novel ideas
are required in the field to find the reasons for such spread of the results.

In the present paper we try to find out possible reasons for the large difference between the results of different experiment.
In order to do it we analyze the thermodynamic properties of graphite and liquid carbon obtained in different experiments and
theoretical models.



In order to find some reference points on the phase diagram we start from the discussion of graphite-liquid-vapor triple and critical
points of carbon. The estimations of the triple point suffer on great uncertainty \cite{savvatimsky}. At the same time
there are just a few evaluations of the critical point of liquid-gas transition in carbon. The most well recognized
paper is the one by Leider, Krikorian and Young \cite{leider}.

In order to evaluate the critical temperature of carbon, the authors of this paper had to make several assumptions.
According to the available for them experimental results the specific volume of liquid carbon had to be substantially
larger than the one of graphite, but they did not have exact values of the density of molten carbon. For this reason
they arbitrary assumed that the volume jump upon melting of carbon is $20 \%$. This assumption gave them the density of liquid
carbon at the triple point $\rho_{liq-tr}=1.611$ $g/ml$.

The second assumption of Ref. \cite{leider} is that thermal expansion coefficient of liquid carbon is the same that the one of
graphite. Using the law of rectilinear diameters the authors estimate the critical point parameters as $T_c=6810$ K, $\rho_c=0.64$ $g/ml$
and $P_c=2200$ atm.

It is seen from the description above that two suspicious assumptions are introduced in Ref. \cite{leider}. For instance, the
later experimental result state that the volume jump of carbon upon graphite melting is much larger: $V_l/V_s=1.7$ (see paragraph 5.2 of
Ref. \cite{savvatimsky}).

\begin{figure}

\includegraphics[width=8cm, height=6cm]{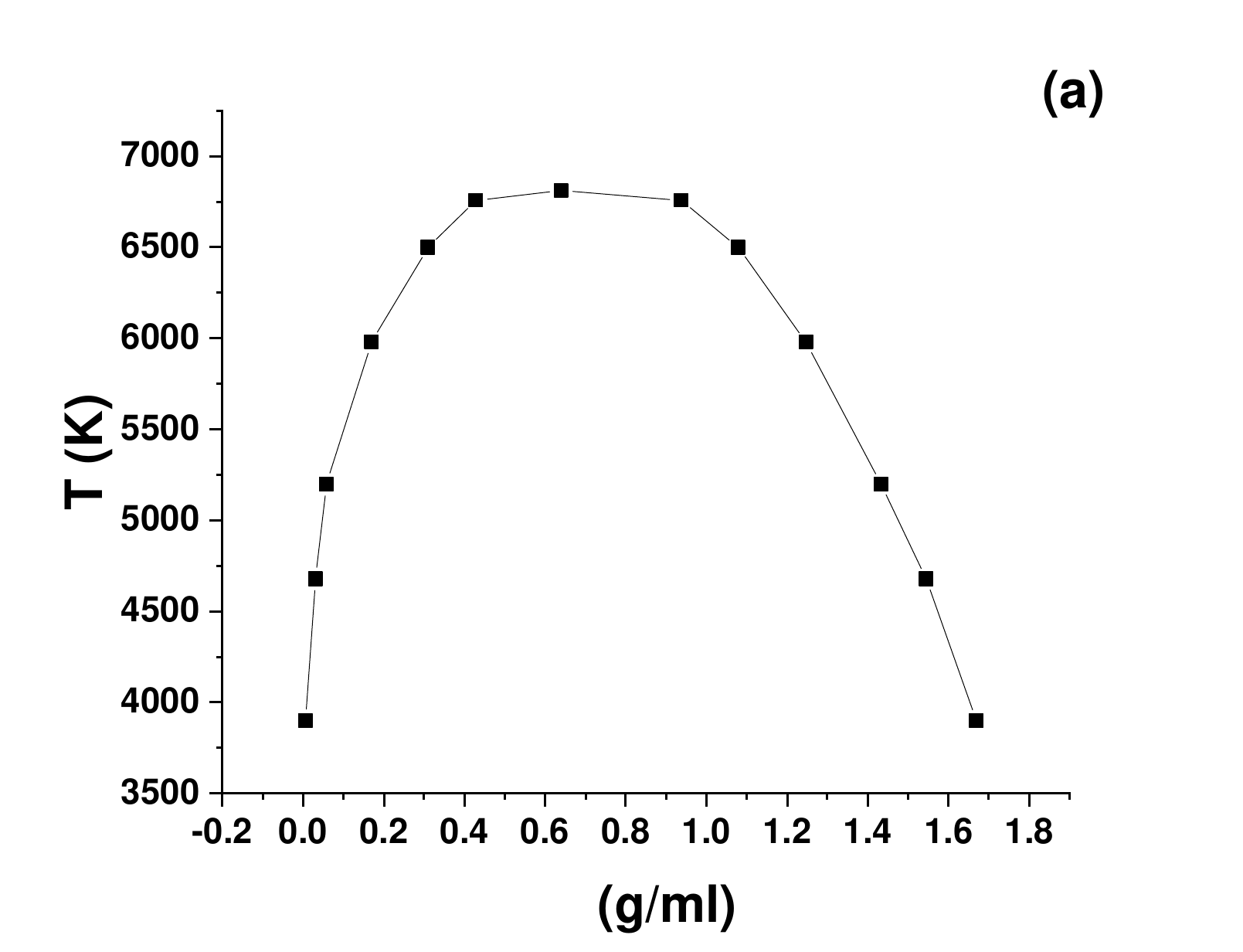}%

\includegraphics[width=8cm, height=6cm]{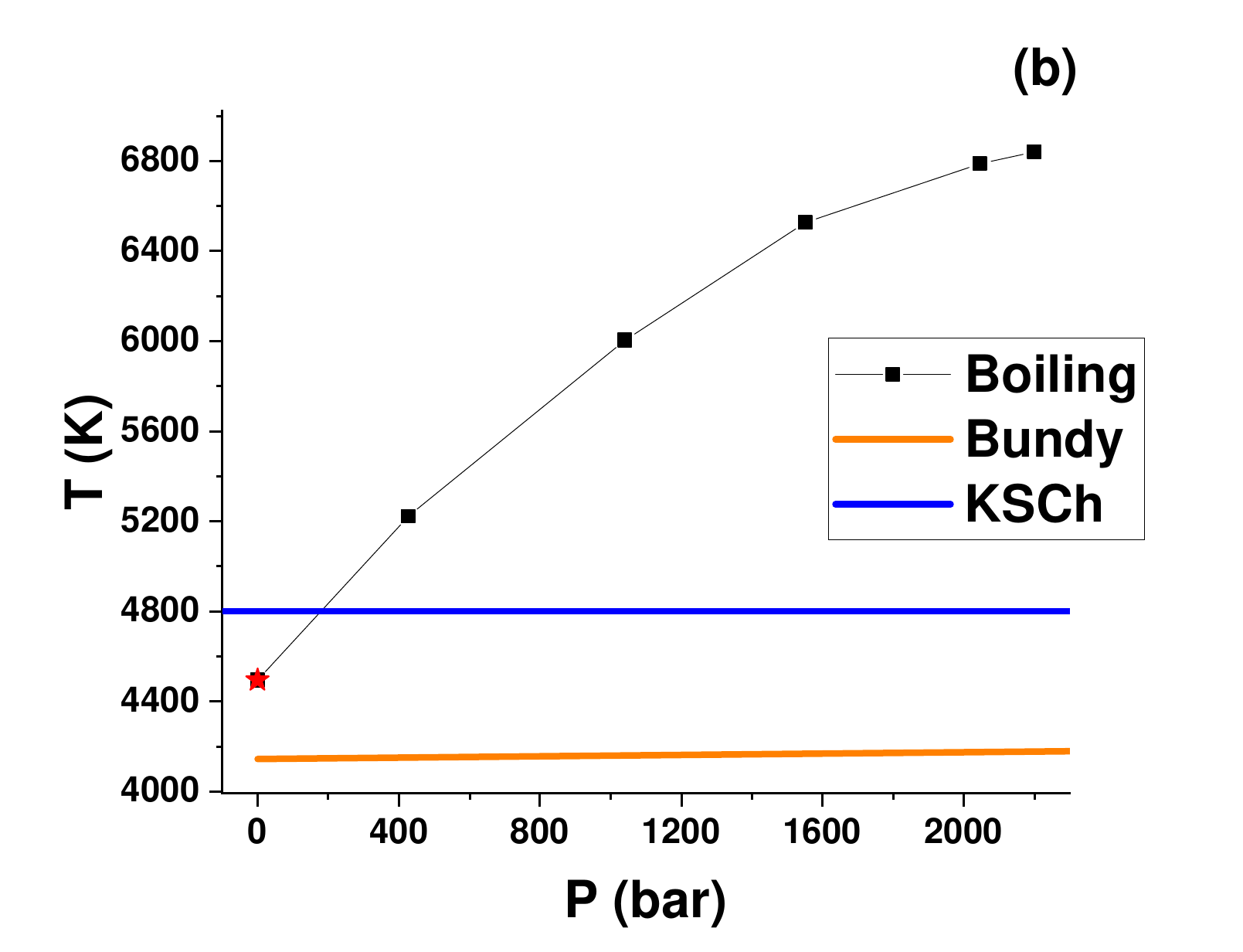}%

\caption{\label{boiling} Boiling curve of carbon obtained by the corresponding states law with the
critical parameters from Ref. \cite{leider} in (a) $\rho - T$ and (b) $P-T$ planes. The red star in the panel (b)
shows an extrapolation of the boiling curve to the limit of zero pressure. The curves 'Bundy' and KSCh in the panel
(b) show the melting curves of graphite from Refs. \cite{bundy} and \cite{ksch} correspondingly.}
\end{figure}

In Fig. \ref{boiling} we show the boiling line of carbon obtained by rescaling the boiling line of Lennard-Jones system according
to the law of corresponding states. The critical parameters of carbon were taken from Ref. \cite{leider}. In the $P-T$ plane
we show also the melting curve of graphite from a seminal work by Bundy \cite{bundy} and the melting curve obtained by
Korobenko, Savvatimskiy and Cheret \cite{ksch}. By extrapolating the curve to the zero pressure limit we obtain that the boiling curve
does not cross the melting line from the Bundy work, i.e. no triple point is identified. If we take the melting curve from Ref.\cite{ksch}
we obtain that the triple point takes place at $P=180$ bar and $T=4800$ K. The triple point obtained in Ref. \cite{leider} is in the range
$4600-4800$ K and $P=103$ bar. A careful analysis of the most of the available data given by Savvatimsky in Ref. \cite{savvatimsky} also gives
the location of the triple point at $P \approx 120$ bar and $T \approx 4800 - 4900$ K.

The analysis of experimental data also requires some parameters which are not known with sufficient accuracy. For instance,
it is supposed in Ref. \cite{ksch} that the heat capacity of graphite prior to the melting point
is $c_P=3.2$ $\frac{J}{g \cdot K}$ for the temperature range from $T=4200$ K to $5200$ K. At the same time
the authors compare their results for the heat capacity with the literature data: $c_P=3.27$ $\frac{J}{g \cdot K}$
at $4400$ K and increases to $4.11$ $\frac{J}{g \cdot K}$ in Ref. \cite{cp-1} and $c_P=3.77$ $\frac{J}{g \cdot K}$ at $T=4500$ K
in Ref. \cite{cp-2}. Although the difference between different data is not so crucial, the amount of
absorbed heat is determined by an integral of the heat capacity, therefore, this disagreement can result
in completely different estimation of the melting point.


Another kind of inconsistencies take place in a number of other works. For instance, in Ref. \cite{rakhel} experimental measurements of
'liquid carbon' density at $T=7000$ K are reported. However, this temperature exceeds the $T_c$ from Ref. \cite{leider}, therefore
no 'liquid', but supercritical fluid takes place in this region. At the same time, measurements of Ref. \cite{rakhel} show
that the isochoric heat capacity of 'liquid' carbon is $c_V/R=3.4 \pm 0.4$. This value looks to be very small for a region of
phase diagram close to the critical point, where the heat capacity demonstrate strong grows. We remind that according to
Ref. \cite{leider} the critical temperature $T_c=6810$ K is the estimation of the upper limit of critical temperature of
carbon. It means that the real critical temperature can be much lower. However, strong maximum of isochoric heat capacity
preserves up to the temperatures of $T \approx 3 T_c$ \cite{widom-lj}, therefore the heat one should still expect large $c_V$.

The examples above demonstrate an important lack in the analysis of the experimental data: although experiments on graphite melting
are performed at extremely high temperatures, in most cases influence of the gas phase is not taken into account. Interestingly,
the authors of experimental papers often mention that the presence of sublimated carbon influence the results of measurements,
for instance, worsening the data of pyrometry, but the presence of gas phase is not taken into account in the thermodynamic
calculations to develop the experimental data. At the same time it looks clear that sublimation of carbon should
influence the heat capacity of the sample.


In the present paper we try to find out the location of the critical point of carbon by  molecular dynamics method. Having determined the
location of the critical point we calculate the points of maxima of thermodynamic response functions (Widom lines) and discuss their influence
on the experimental measurements of carbon melting.



\section{System and Methods}

In the present work we simulate a system of 4096 carbon atoms in a cubic box with periodic boundary conditions.
GAP-20 potential \cite{gap20} is used to model the interactions between the particles. The same model was used
in Ref. \cite{or-gap} for simulation of fluid state of carbon at high temperature. The system was modeled in
canonical ensemble (constant number of particles N, volume V and temperature T). The timestep was set to
$dt=2.0$ fs. The equilibration period was $10$ ps. The production period was set to $20$ ps.

Since we are interested in liquid-gas critical point we had to simulate a system in a large number of state points.
The density varied from $\rho_{min}=0.05$ $g/ml$ to $\rho_{max}=2.0$ $g/ml$ with step $\Delta \rho=0.05$ $g/ml$.
The temperatures varied from $T=4000$ K to $5000$ K with step $\Delta T=200$ K. Additionally we calculated the
$T=4100$, $6000$ and $7000$ K isotherms.

During the course of simulation we sampled the thermodynamic quantities such as pressure and internal energy. We constructed
the isotherms of the system and identified the critical point as a point of inflection on the isotherm.
We approximate the thermodynamic data from simulation by an equation $X=\sum_{i,j=0,4}a_{ij}T^i\rho^j$,
where $X$ stands for internal energy or pressure and $a_{i,j}$ are fitting coefficients. In this approximation
we used only the data for $\rho \leq 1.0$ $g/ml$. This density interval was sufficient for investigation of near critical
anomalies. The fitting coefficients are given
in Supplementary materials. Using these fits we calculated the thermodynamic response functions such as isobaric
heat capacity $c_P=\left( \frac{\partial H}{\partial T} \right)_P$, where $H=U+PV$ is the enthalpy of the system,
thermal expansion coefficient $\alpha_P=\frac{1}{\rho} \left( \frac{\partial \rho}{\partial T} \right)_P$ and
isothermal compressibility $\beta_T=\frac{1}{\rho} \left( \frac{\partial \rho}{\partial P} \right)_T$. We calculated the
location of the maxima of these quantities.

All simulations were performed using the LAMMPS simulation package \cite{lammps}.

\section{Results and Discussion}

\begin{figure}

\includegraphics[width=8cm, height=6cm]{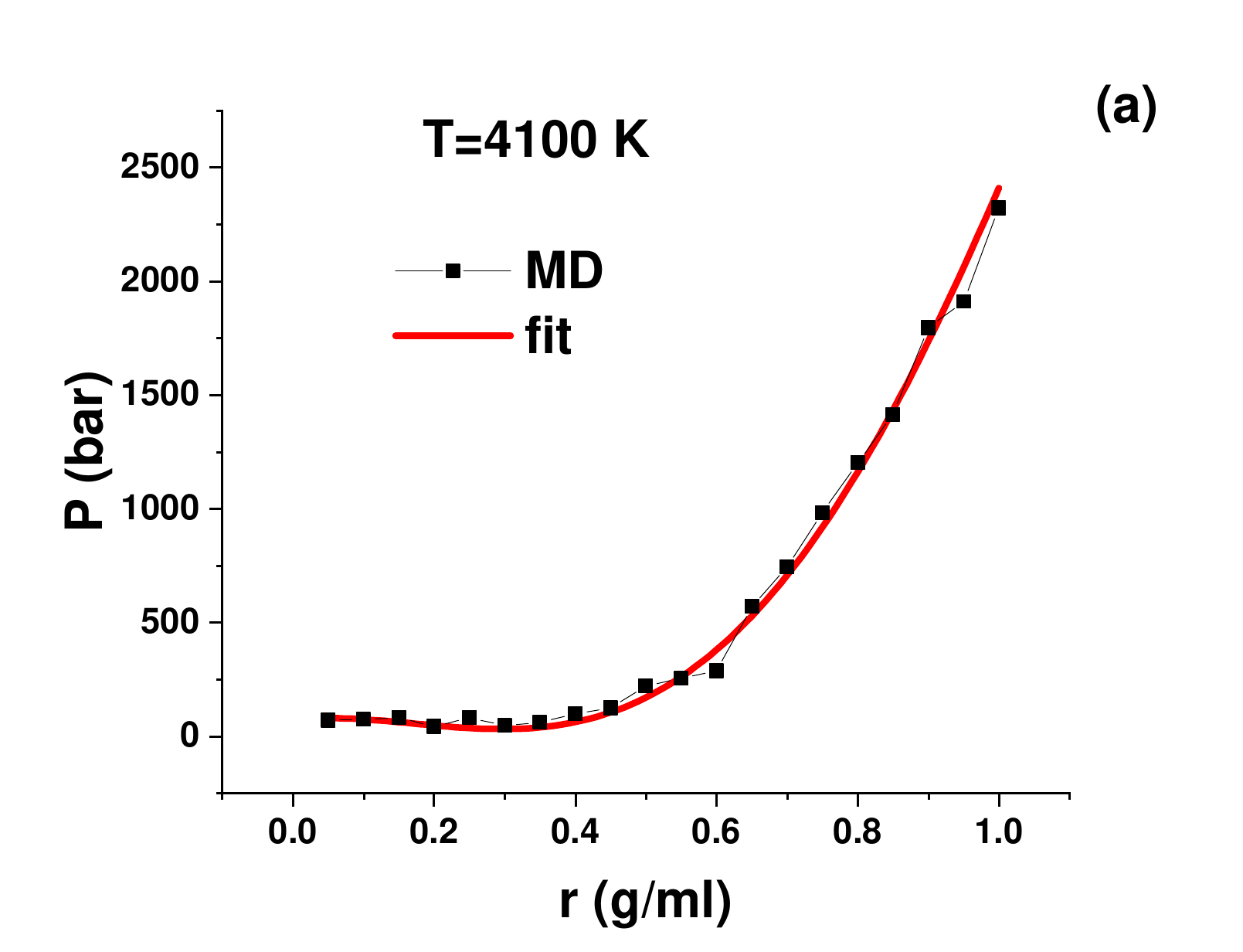}%

\includegraphics[width=8cm, height=6cm]{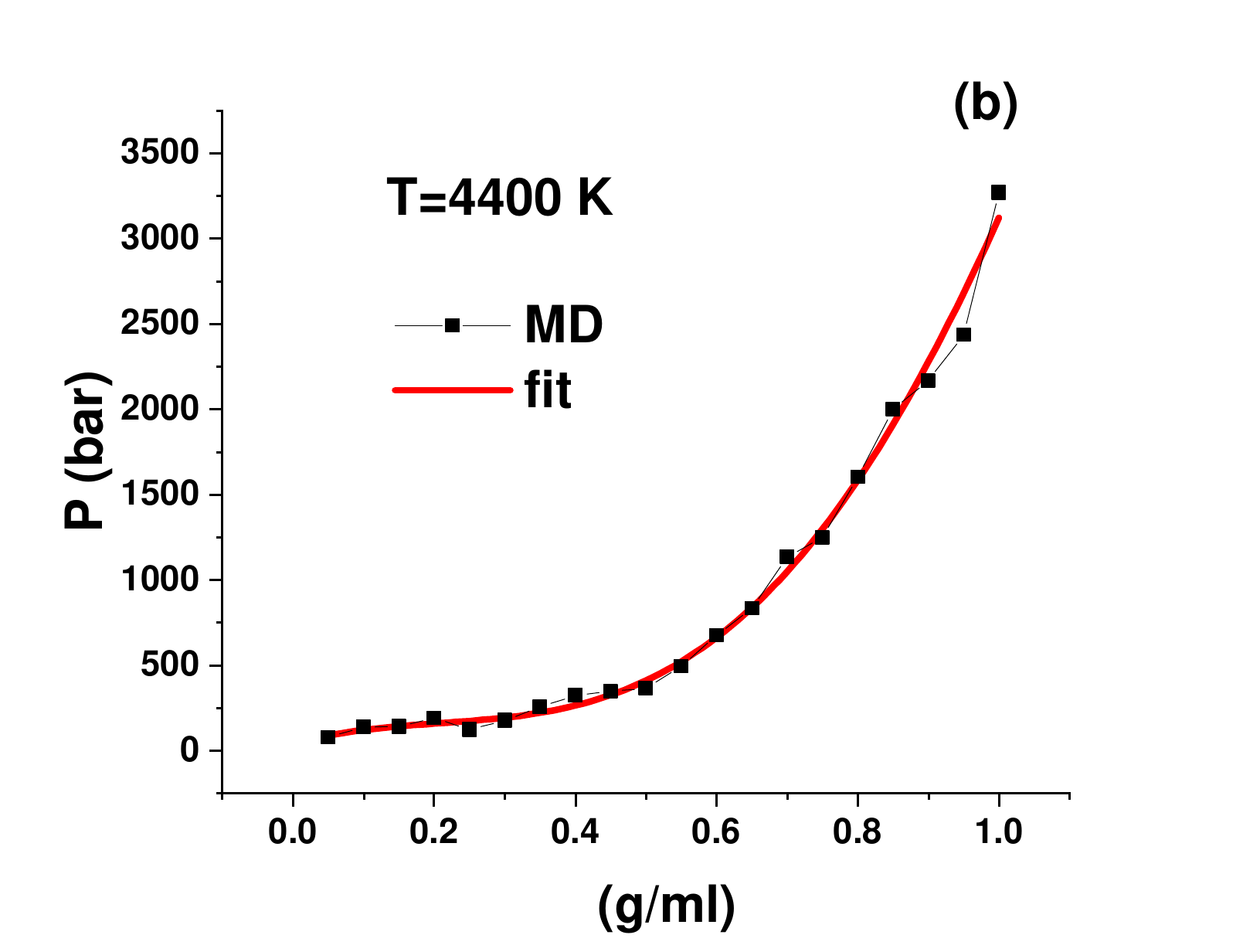}%

\includegraphics[width=8cm, height=6cm]{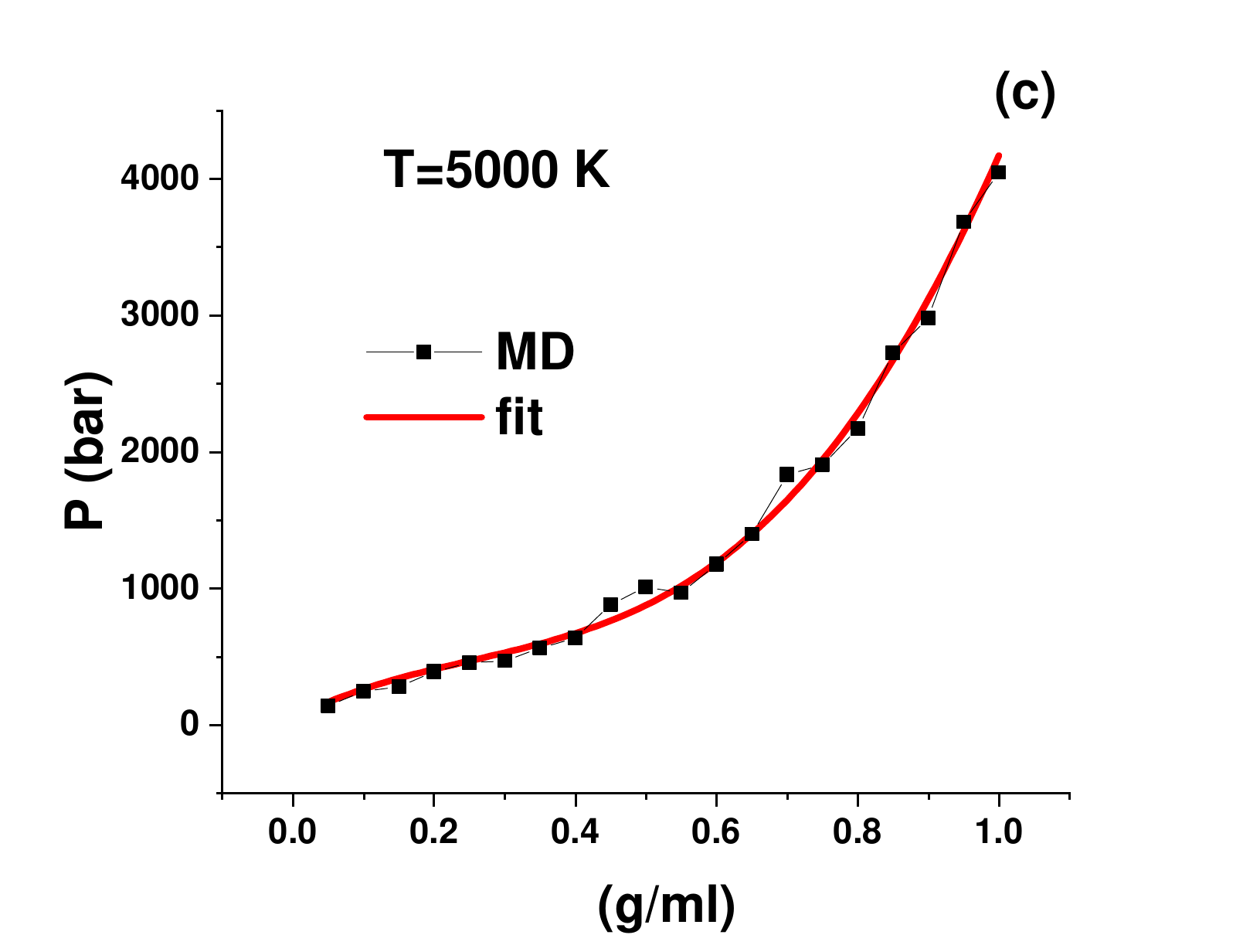}%

\caption{\label{fit} Equation of state and fitting curve of GAP-20 model of carbon at (a) $T=4100$ K, (b) $T=4400$ K
and (c) $T=5000$ K.}
\end{figure}

First of all we calculate the equation of state (EoS) of carbon and fit it to the polynomial function described above.
The examples of EoS along isotherms and fit of the data are shown in Fig. \ref{fit} (a) - (c). The MD data are rather
noisy. The GAP-20 potential is rather computationally heavy. For this reason, the MD simulation was relatively short
which did not allow to collect very good statistics. The fitting curves are in good agreement with the results of simulation.
Therefore, we assume that one can use the fitting polynomial to determine the thermodynamic properties of carbon.

\begin{figure}

\includegraphics[width=8cm, height=6cm]{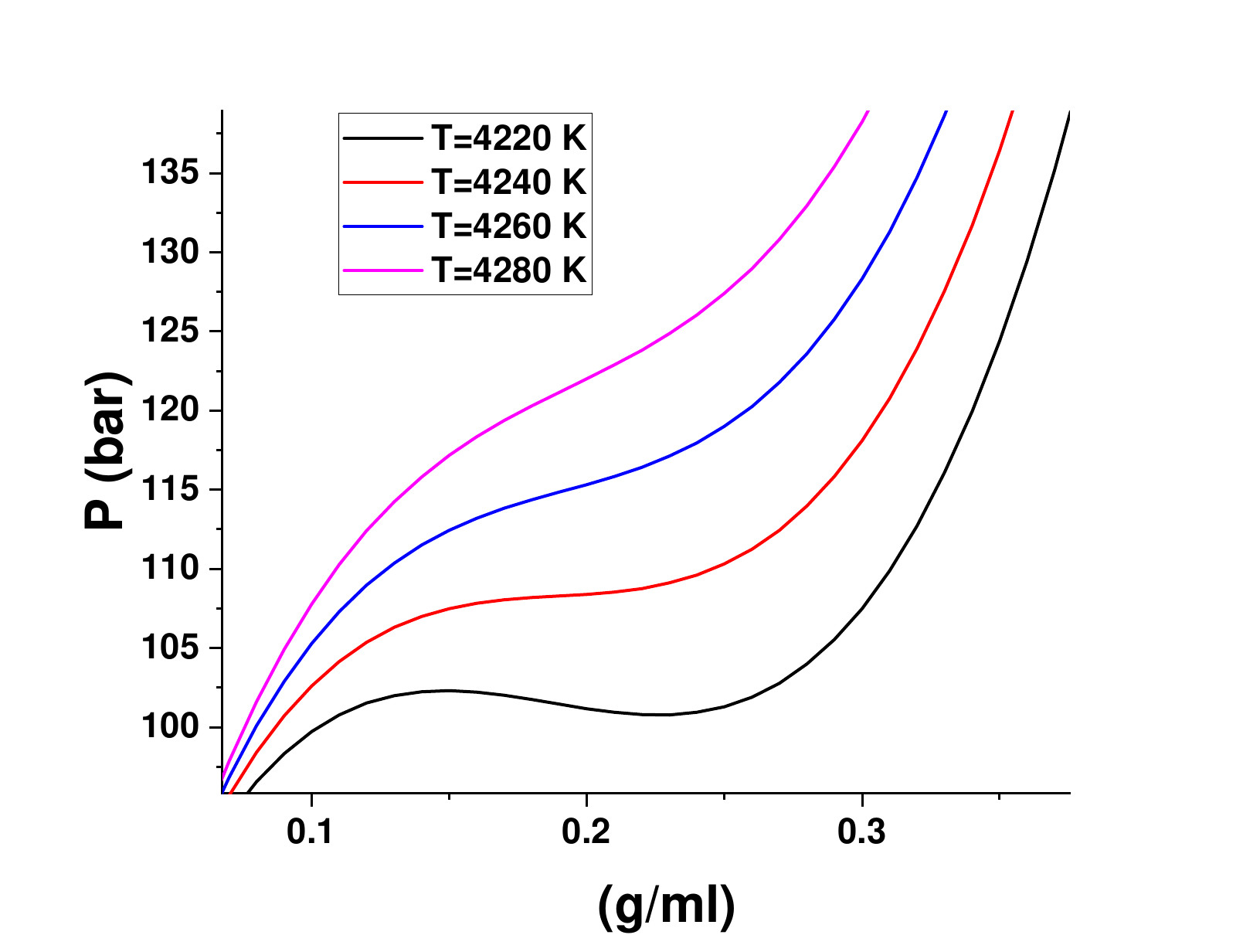}%

\caption{\label{tc} Several isotherms next to the critical temperature of carbon obtained from the polynomial fit.}
\end{figure}

Figure \ref{tc} shows several isotherms in the vicinity of the critical temperature. From these isotherms we estimate the critical
point as $T_c=4235$ K, $\rho_c=0.1887$ $g/ml$ and $P_c=106.61$ bar. It is much lower than the one from Ref. \cite{leider}. The estimation
of graphite-liquid-gas triple point in the same paper is $T_t=4765$ K, which is above our estimation of $T_c$. At the same time,
experimental evaluation of triple point of carbon suggests the values from $T_t=3670$ K to $4650$ K (see table 3.1 in Ref. \cite{savvatimsky}),
therefore $T_c$ of our simulations can be consistent with some experimental works. Any way, we would like to stress that evaluation of the
critical point is a difficult task which requires special care. Even tiny mistakes in interaction potential can lead to strong
deviation of calculated critical point from the experimental one. The GAP-20 model has been fitted to simulate condensed phases
of carbon and it can be not good enough for the critical point evaluation. However, this result is valuable, since it demonstrates
that accounting for the critical phenomena is required in analysis of experimental data on graphite melting.

At the same time at $T=4100$ K and $\rho=1.55$ $g/ml$ we observe that system demonstrates some traces of crystallinity. It means that the critical point
is located rather close to the melting line in the density-temperature plane. It is well known that near the critical point some quantities
(heat capacity, thermal expansion coefficient, etc.) demonstrate maxima (Widom lines). This case when the melting process is complete
the system can approach the lines of maxima of these quantities, which should strongly influence the thermodynamic response of the
system, for instance, the heat absorption of molten carbon. We remind that in many publications the heat capacity of the liquid
carbon is assumed to be constant in a wide range of temperatures. This assumption is violated in the vicinity of the critical point.
Moreover, since the thermodynamic function demonstrate strong dependence on pressure and temperature, even small uncertainty
in these quantities can lead to a large discrepancy of the results, which can be responsible for a large body of conflicting measurements
of the carbon melting point.

\begin{figure}

\includegraphics[width=8cm, height=8cm]{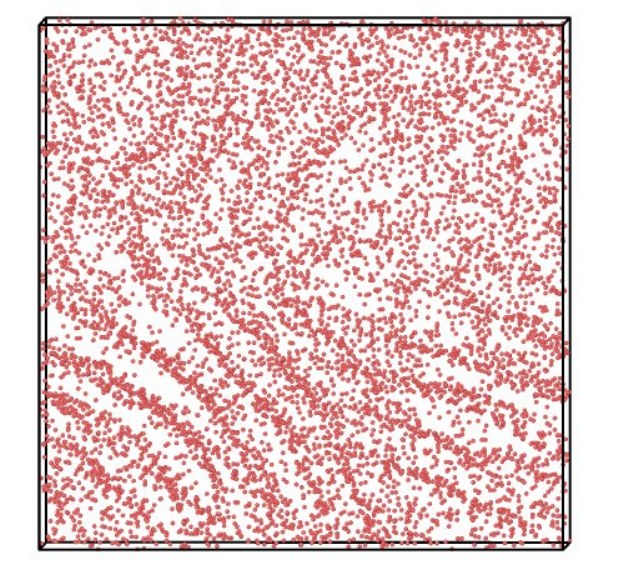}%

\caption{\label{cryst} A snaphot of a sample at $T=4100$ K and $\rho=1.55$ $g/ml$. Traces of crystallization
are visible in the bottom part of the Figure.}
\end{figure}

\begin{figure}

\includegraphics[width=8cm, height=6cm]{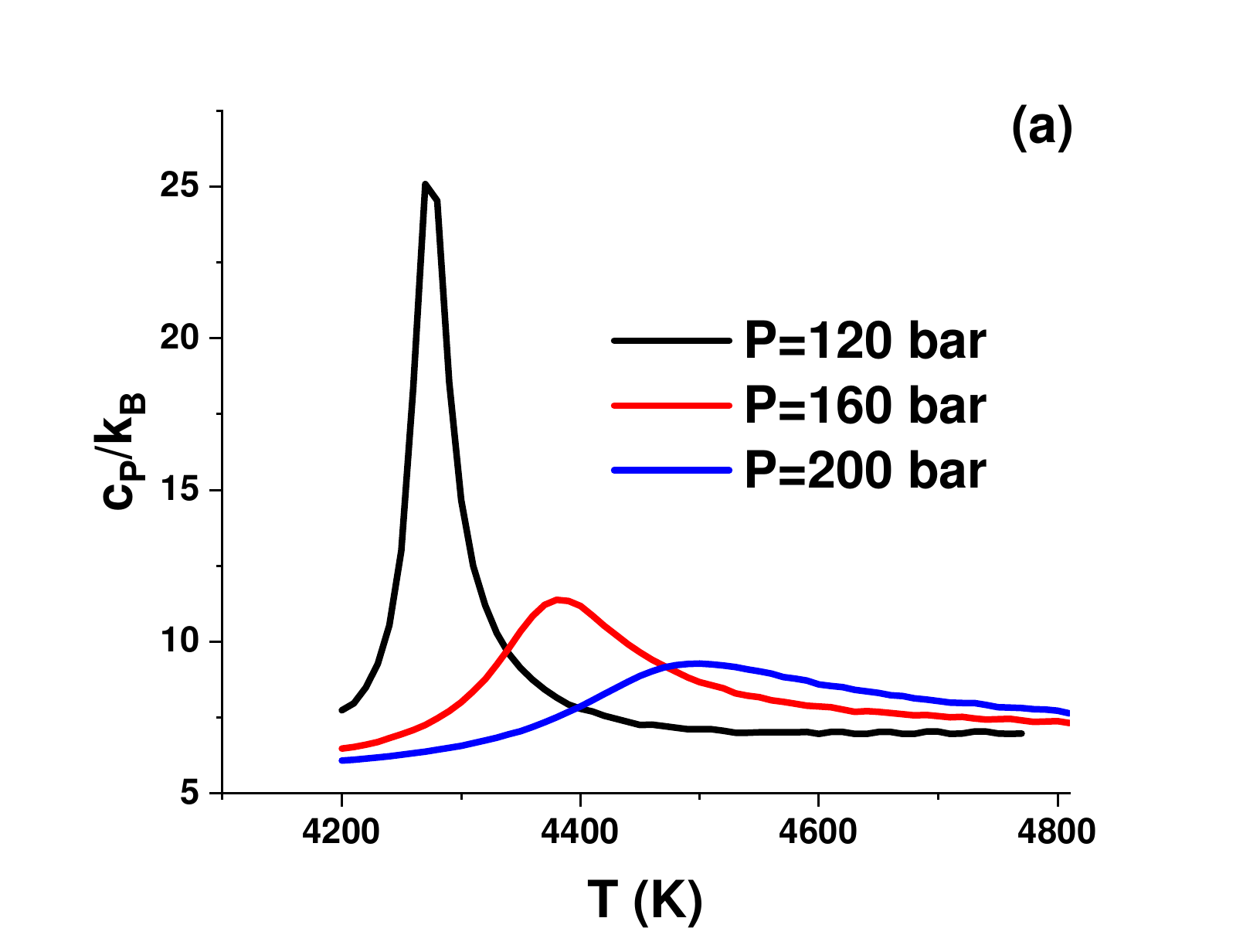}%

\includegraphics[width=8cm, height=6cm]{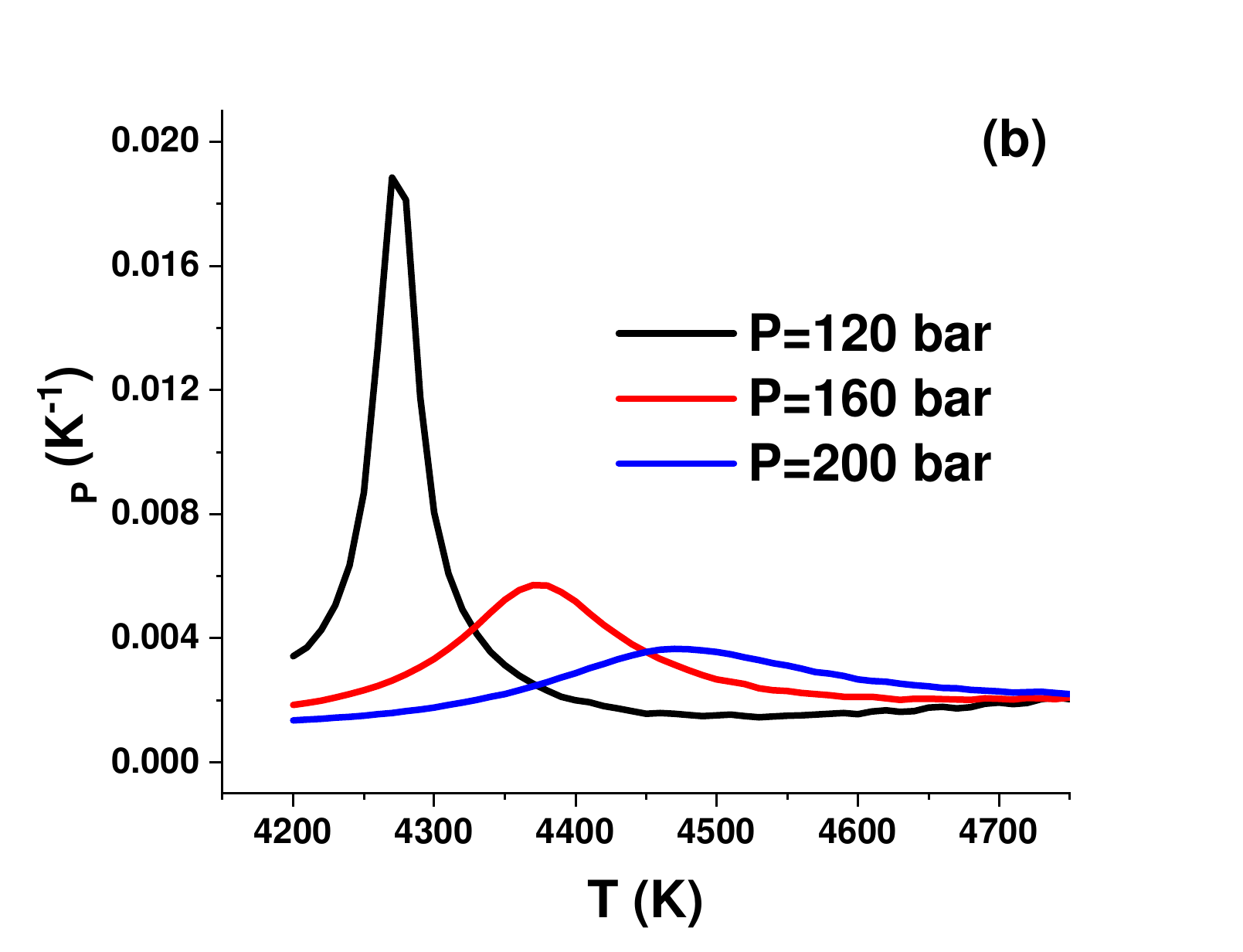}%

\includegraphics[width=8cm, height=6cm]{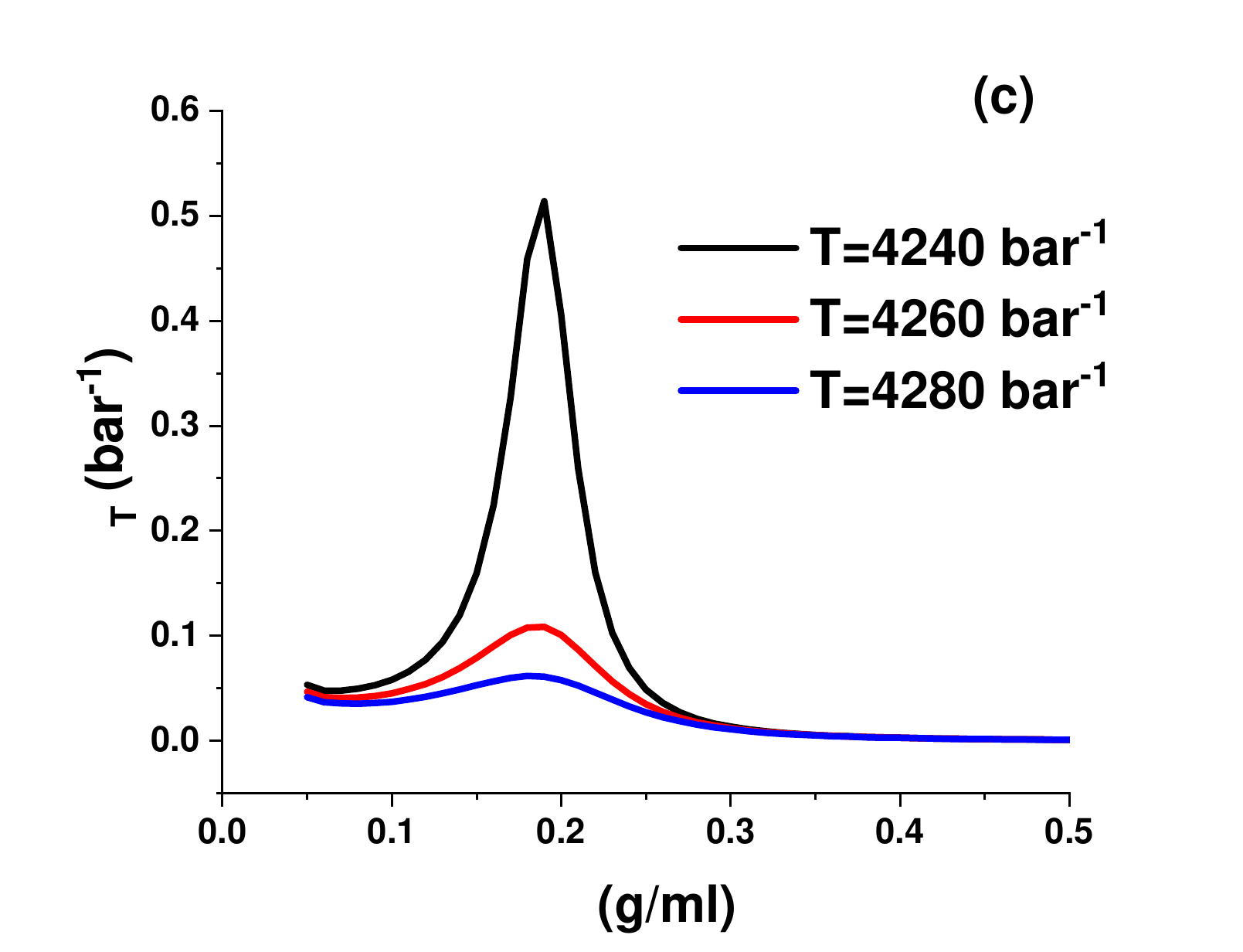}%

\caption{\label{max} (a) Isobaric heat capacity, (b) thermal expansion coefficient and (c) isothermal compressibility
of GAP-20 model of carbon next the critical point.}
\end{figure}

Based on the discussion above it is impotant to calculate the near-critical maxima of different thermodynamic functions of GAP-20 carbon.
Examples of isobaric heat capacity, thermal expansion coefficient and isothermal compressibility of GAP-20 carbon are given
in Fig. \ref{max}. As it is expected, all these quantities demonstrate maxima close to the critical point, which rapidly disappear
as we move away from it.

We recall that in Ref. \cite{rakhel} large heat capacity of liquid carbon at $T=7000$ K was reported: $c_V/R=3.8 \pm 0.4$. Later on
in Ref. \cite{or-gap} an attempt to explain the high value of $c_V$ was undertaken. The authors assumed that the reason for
large $c_V$ can be due to transition from high density sp2 hybridized liquid to an sp hybridized low density one. They also
observe voids in the 'liquid' and negative slope of the density dependence of isothermal expansion coefficient $\beta_T$. All these
results are compatible with the assumption that 'liquid' carbon is indeed in supercritical state (below we use the abbreviation SCF
for 'supercritical fluid'). While  gaseous carbon
is mostly sp hybridized, the liquid one has sp2 hybridization. A smooth crossover from gas-like to liquid-like state
takes place at supercritical temperatures. SCF demonstrates clusters at low densities and voids at higher one. Finally,
the region of densities studied in Ref. \cite{or-gap} belongs to decreasing branch of isothermal compressibility, whose maximum
takes place at the densities about $\rho=0.2$ $g/ml$ (see Fig. \ref{max} (c)).

\begin{figure}

\includegraphics[width=8cm, height=6cm]{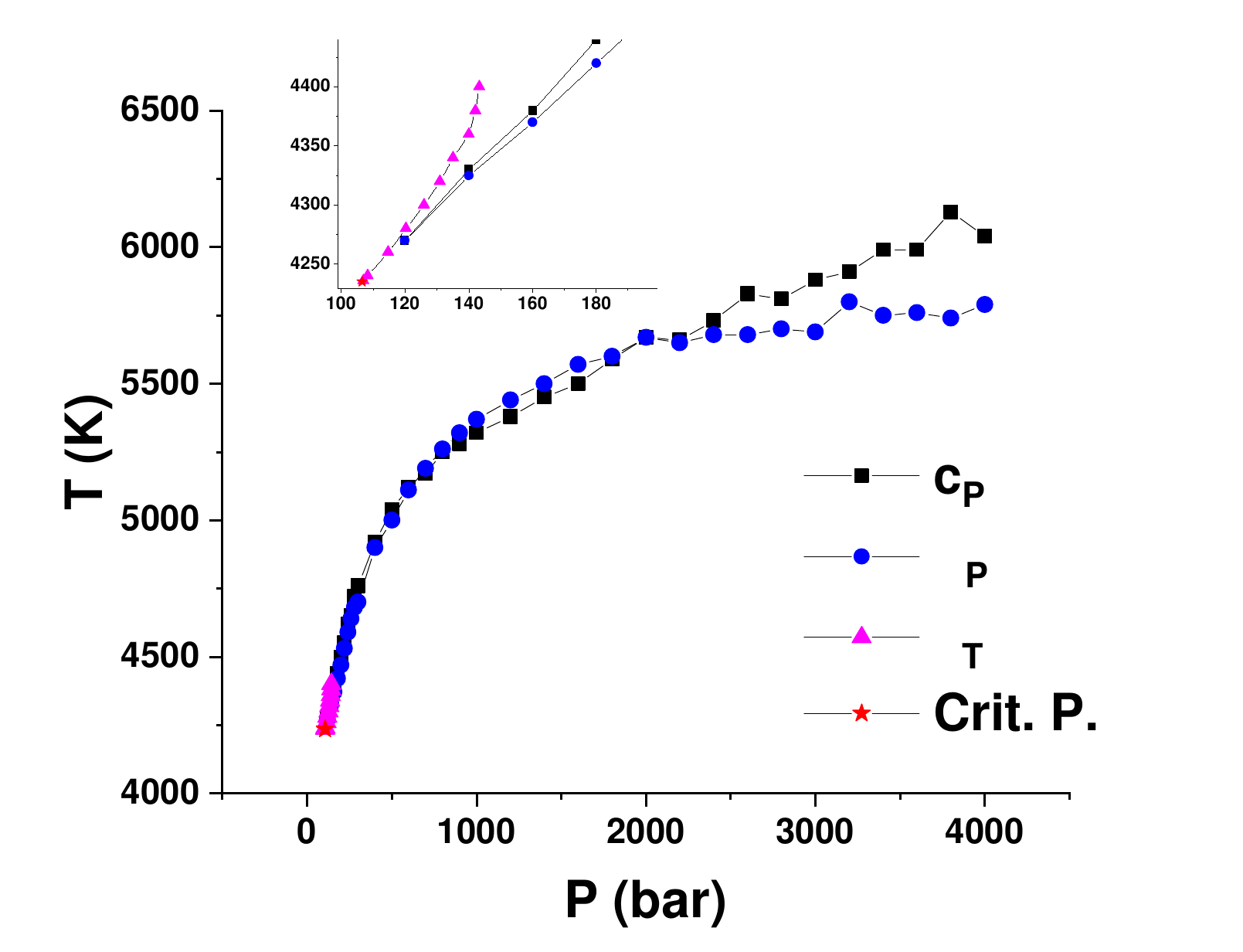}%

\caption{\label{widom} Location of supercritical maxima of GAP-20 model of carbon.}
\end{figure}

The locations of maxima of thermodynamic response functions of GAP-20 model of carbon are shown in Fig. \ref{widom}.
This figure looks qualitatively similar to the Widom lines of other substances: van der Waals gas \cite{vdw}, Lennard-Jones system \cite{lj-wid},
square well \cite{sw-wid}, carbon dioxide \cite{co2-wid}, etc. The maxima of $\beta_T$ take place only at the temperatures slightly
exceeding the critical one and rapidly disappear. At the same time the maxima of $c_P$ and $\alpha_P$ preserve
even at rather large temperatures. We recall that in the case of simple liquid these maxima take place up to the
temperatures about $3T_c$. We have not done the simulations at such high temperatures.

Calculation of location of triple point of carbon requires estimation of both melting and boiling lines, which goes beyond
the scope of the present paper. In our simulation we observe spontaneous crystallization of liquid carbon
into graphite phase at $T=4100$ K, $\rho_{gr}=1.55$ $g/ml$, $P_{gr}=7225$ bar and $T=4200$ K, $\rho_gr=1.8$ $g/ml$,
$P_{gr}=20262$ bar. These values are close to the critical point in temperature, but exceed it in pressure. However,
the density rapidly decreases upon isobaric heating which can result in approaching of the Widom line in the
experimental studies of liquid carbon.

It is of special interest to have a look on the maxima of isobaric heat capacity and thermal expansion coefficient (Fig. \ref{max} (a) and (b)).
It is seen, that in the close vicinity of the critical point the isobaric heat capacity reaches as much as $25k_B$ per particle.
It rapidly drops on moving away from the critical point. Any way, it is clear that the assumptions of constant heat capacity
completely fail.

The same speculation takes place for the thermal expansion coefficient, which is supposed to be close to the one of graphite:
in the vicinity of the critical point $\alpha_P$ is much larger than it would be expected from the one of graphite.

We would like to stress once again: the results of this work are obtained with the help of GAP-20 model, which was developed
to describe condensed phases of carbon. For this reason, these results cannot be directly projected on the experimental data
and can be used for qualitative speculations only. However, such speculations allow us to propose a possible explanation
of large spread of experimental data on melting point of graphite.

The melting temperature of graphite appears to be very close to the critical one. As a result when the crystal structure collapses,
the system appears to be in supercritical state with strong fluctuations of thermodynamic quantities (see Fig. \ref{widom}).
These fluctuations play two negative roles in the analysis of the experimental data. First of all, to the best of our knowledge,
they are not accounting for when the raw data are analyzed. The second point is that even small experimental uncertainties
lead to great errors. For instance, when the system is close to the maximum of isothermal compressibility even small
error in pressure leads to strong error in density.

\section{Problems of the critical point calculations}

In the present paper we discuss calculation of critical point of carbon. The results obtained in our calculations
look very odd, which is, however, consistent with general understanding of the critical point calculations.


The GAP-20 potential employed in the present study is a machine-learning model fitted to reproduce the DFT calculations. Therefore, it
cannot be more accurate than the DFT calculations themselves. However, calculation of critical point of gas-liquid transition
is a challenging problem for DFT and indeed the DFT results are not trustable themselves.

Several problems take place to calculate the location critical point in DFT. First of all, the critical point of carbon is
located at very high temperature, which looks unreasonably high for DFT. Also van der Waals (vdW) forces are of great importance in
the vicinity of the critical point. The optB88-vdW functional used for fitting the GAP-20 model employees the van der Waals
corrections, which are important for the reliable graphite calculations. This functional shows good results for solids
\cite{dft-1,dft-2} and surfaces of crystals \cite{dft-2}. However, it was not tested for diluted phases such as gaseous carbon.
Therefore, large errors can be expected for gases and supercritical fluids.

More problems take place in the vicinity of the critical point, such critical fluctuations and critical slowing down which i) require to
simulate rather large system and ii) require to perform very long simulations to get a lot of data for proper averaging. Although
GAP-20 potential works much faster than ab-initio calculations, we are still unable to overcome these problems (the system size
is 4096 atoms and the production run is 20 ps). Moreover, since large fluctuations take place next to the critical point,
even tiny error in some parameters strongly affects the others. For instance, the well recognized model of water TIP4P/2005
gives the critical temperature $T_c=640$ K (the experimental value is $T_c^{exp}=647$ K), critical density
$\rho_c=0.31$ $g/ml$ (the experimental one $\rho_c^{exp}=0.322$ $g/ml$), while the critical pressure is
$P_c=146$ bar ($P_c^{exp}=220.64$ bar) \cite{tip4p}, i.e., a relatively small error in the critical density
leads to a great effect on critical pressure: $P_c^{exp}/P_c=1.5$. This example shows that even in the case
of well developed systems with numerous experimental data available the evaluation of the critical point
parameters is still challenging.

In the case of substances with low critical temperature the location of the critical point is used as
one of the fitting parameters. For instance, Lennard-Jones potential parameters of noble gases can be selected
in such a way that the critical point of the model corresponds to the experimental critical point of the substance of
interest. This is, however, not applicable to the case of carbon.

\bigskip

Based on this discussion, we are sure that the location of the critical point reported in the present study does not
correspond to the critical point of real carbon. At the same time we believe that the calculation of the present paper
make sense. These calculations let us to propose that one of two possibilities can take place in the case of experimental
investigations of melting of graphite.

The first option is the one obtained in our simulations: after the heating of graphite the system transforms into SCF.
This case it experiences strong fluctuations, including maxima of heat capacity (which influences the estimation of the
temperature) and thermal expansion coefficient (which leads to wrong evaluation of the density).

The second option which should take place if the critical point is at the temperatures above the ones reported in
the melting experiments is that after the melting of graphite the system can be into the liquid-gas two phase region.
However, the system in two phase region is also very sensitive to tiny perturbations, since they change the
amount of particles in the phases.

In both cases one should be very careful to the thermodynamic properties of the system and the usual assumptions
like constant heat capacity or constant thermal expansion coefficient should give erroneous results.

Novel measurements of the graphite melting appeared recently \cite{rakhel-new}. According to the results of this paper,
the melting temperature of graphite in the pressure range of 0.2-2.5 GPa is about 6000 K and demonstrate very
weak dependence on pressure. The estimation of the volume of the solid and liquid carbon at the melting point
is $V_s/V_0 \approx 1.3$ and $V_l/V_0 \approx 1.8$, where $V_0=0.4425$ $ml/g$ is the volume of HOPG at ambient conditions.
It gives $V_l/V_s \approx 1.38$, which is different from the values used in \cite{leider} to evaluate the critical point.
The authors estimate the isochoric heat capacity of liquid carbon next to the melting line to be $c_V=2.9$ $J/(g \cdot K)$,
which corresponds to $c_V/k_B \approx 4.19$. This value is much larger than the one corresponding to Dulong-Petit law,
which can happen either in the near-critical fluctuations region, or if some structural crossovers in the liquid
phase take place \cite{highcv-1,highcv-2}. At the same time the authors write that the thermal expansion coefficient is nearly constant
in the temperature range of 4000-6000 K, which is not consistent with the assumptions above. Once again
we see that even in the case of careful measurements no self-consistent result is achieved.

The authors of Ref. \cite{rakhel-new} also compare their results with the ones of the quantum molecular dynamics
simulations. The simulations give the melting temperatures of graphite of the order of 3800 K. The GAP potential employed in the
present study should give the melting point of the same order of magnitude. Comparing different experimental and computational
works one can find that simulations give too low melting points comparing to the experiments. One can assume that
some drawbacks of quantum simulations lead to underestimation of the melting points. Further development of the
simulation methods is definitely required.

\section{Conclusions}

In the present paper we perform molecular dynamics simulation of carbon with GAP-20 machine learning
potential. We show that this model demonstrates a very low critical temperature of gas-liquid transition:
$T_c=4235$ K, which is very close to the results for the melting temperature of graphite. For the first
time we calculate the Widom lines of carbon in the framework of GAP-20 model. Based on these results we propose
a possible mechanism of inconsistencies between different experiments on graphite melting:
they might be related to the strong fluctuations of thermodynamic quantities in the
vicinity of the critical point.

We note that the results are obtained in the framework of GAP-20 model, which was constructed for condensed phases
of carbon. For this reason the results are not applicable to direct comparison with experiment. At the same time
comparing different computational and experimental works, one can assume that the simulations give
qualitatively correct results, but shifted with respect to the experimental ones to lower temperatures.
This case if can happen that the idea of the present paper is correct, but the critical point is at much higher temperatures.
This assumption is consistent with the critical temperature of carbon from Ref. \cite{leider} ($T_c=6810$ K) and
the results for the melting temperature of graphite obtained in the recent work \cite{rakhel-new} $T_m \approx 6100$ K.

We believe that construction of novel models of carbon, fitted to  the properties of liquid, gas and graphite phases
at the same time is required for more elaborate simulation of the system.





\section{Acknowledgments}

I acknowledge V.V. Brazhkin (HPPI RAS) and N.M. Chtchelkatchev (JINR) for stimulating discussions.
This work has been carried out using computing resources of the federal
collective usage center Complex for Simulation and Data Processing for
Mega-science Facilities at NRC "Kurchatov Institute", http://ckp.nrcki.ru/.
The work was supported by the Russian Science Foundation (Grant 25-22-00876).


\begin{thebibliography}{62}

\bibitem{1911} O. P. Watts, C. E. Mendenhall, On the fusion of carbon, Phys. Rev. (seriae 1), 33, 65-69 (1911)




\bibitem{savvatimsky} A. Savvatimskiy, Carbon at high temperatures, Springer Series in Materials Science 134,
Springer International Publishing Switzerland 2015

\bibitem{rakhel} A. M. Kondratyev and A. D. Rakhel, Melting Line of Graphite, Phys. Rev. Lett. 122, 175702 (2019)

\bibitem{leider} H. R. Leider, O. H. Krikorian and D. A. Young, Carbon 11, 555-563 (1973).

\bibitem{bundy} F. P. Bundy, Melting of Graphite at Very High Pressure, J. Chem. Phys. 38, 618 (1963).

\bibitem{ksch} V. N. Korobenko, A. I. Savvatimski, and R. Cheret, Graphite Melting and Properties of Liquid Carbon,
International Journal of Thermophysics, 20, 1247 (1999).


\bibitem{cp-1} G. A. Bergman, L. M. Butchnev, I. I. Petrova, V. N. Senchenko, L. R. Fokin, V. Ya.
Chekhovskoi, and M. A. Sheindlin, in Tahlitsy standartnykh spravochnykh dannykh (State
Service of Standard Referee Data), Grafit kvazimonokristallicheskii, Moscow, GSSD
25-90 (1991).

\bibitem{cp-2} M. A. Sheindlin and V. N. Senchenko, Doklady Akad. Nauk (Fizika) 298, 1383 (1988)


\bibitem{widom-lj} V. V. Brazhkin, Yu. D. Fomin, A. G. Lyapin, V. N. Ryzhov, and E. N. Tsiok,
J. Phys. Chem. B 115, 14112-14115 (2011)

\bibitem{s-liq} A. I. Savvatimskii, S. V. Onufriev, Investigation of the physical properties of carbon
under high temperatures (experimental studies), Physics Uspekhi 63 (10) 1015 - 1036 (2020)

\bibitem{carbon} Yu. D. Fomin, V. V. Brazhkin, Comparative study of melting of graphite and graphene, Carbon 157 (2020) 767-778

\bibitem{gap20} P. Rowe, V. L. Deringer, P. Gasparotto , G. Csanyi, and A. Michaelides,
J. Chem. Phys. 153, 034702 (2020); https://doi.org/10.1063/5.0005084

\bibitem{or-gap} M. Logunov, and N. Orekhov, Carbon 192, 179-186 (2022)


\bibitem{lammps} A. Thompson et. al., Computer Physics Communications 271, 108171 (2022)


\bibitem{vdw} V. V. Brazhkin, V. N. Ryzhov, J. Chem. Phys. 135, 084503 (2011)

\bibitem{lj-wid} V. V. Brazhkin, Yu. D. Fomin, A. G. Lyapin, V. N. Ryzhov, and E. N. Tsiok,
J. Phys. Chem. B 115, 14112-14115 (2011)

\bibitem{sw-wid} V. V. Brazhkin, Yu. D. Fomin, V. N. Ryzhov, E. E. Tareyeva, and E. N. Tsiok,
Phys. Rev. E 89, 042136 (2014)

\bibitem{co2-wid} Yu. D. Fomin, V. N. Ryzhov, E. N. Tsiok, V. V. Brazhkin, Phys. Rev. E 91, 022111 (2015)






\bibitem{dft-1} J. Park, B. D, Yu, S. Hong, Current Applied Physics 15, 885-891 (2015)

\bibitem{dft-2} J. Avelar, A. Bruix, J. Garza, R. Vargas, J. Phys. Condens. Matter. 7; 31, 315501 (2019)
doi: 10.1088/1361-648X/ab18ea.

\bibitem{tip4p} C. Vega, J. L. Abascal, and I. Nezbeda, "Vapor-liquid
equilibria from the triple point up to the critical point for
the new generation of TIP4P-like models: TIP4P/Ew,
TIP4P/2005, and TIP4P/ice", J. Chem. Phys. 125,
034503 (2006)

\bibitem{rakhel-new} A. V. Ivanov, A. M. Kondratyev, and A. D. Rakhel, Phys. Rev. B 113, 014117 (2026)

\bibitem{highcv-1} Yu. D. Fomin, Phys. Chem. Liq. 57, 67-74 (2018)

\bibitem{highcv-2} Yu. D. Fomin, Mol. Phys. 117, 2786-2792 (2019)


\end{thebibliography}
\end{document}